\documentclass[%
 aip,
 amsmath,amssymb,
 reprint,%
]{revtex4-1}

\usepackage{graphicx}
\usepackage{dcolumn}
\usepackage{bm}

\usepackage[utf8]{inputenc}
\usepackage[T1]{fontenc}
\usepackage{mathptmx}

\begin{document}

\preprint{AIP/123-QED}

\title{Quantum transport in high-quality shallow InSb quantum wells}

\author{Zijin Lei$^{\dagger}$}
\email{zilei@phys.ethz.ch}

\author{Christian A. Lehner$^{\dagger}$}
\author{Erik Cheah}
\author{Matija Karalic}

\author{Christopher Mittag}

\author{Luca Alt}

\author{Jan Scharnetzky}

\author{Werner Wegscheider}

\author{Thomas Ihn}

\author{Klaus Ensslin}
\affiliation{ 
Solid State Physics Laboratory, Department of Physics, ETH Zurich, 8093 Zurich, Switzerland
}%

\date{\today}


\begin{abstract}
 $^\dagger$ These authors contributed equally to this work.
~\\
\noindent InSb is one of the promising candidates to realize a topological state through proximity induced superconductivity in a material with strong spin-orbit interactions. In two-dimensional systems, thin barriers are needed to allow strong coupling between superconductors and semiconductors. However, it is still challenging to obtain a high-quality InSb two-dimensional electron gas in quantum wells close to the surface. Here we report on a molecular beam epitaxy grown heterostructure of InSb quantum wells with substrate-side Si-doping and ultra-thin InAlSb (5 nm, 25 nm, and 50 nm) barriers to the surface.  We demonstrate that the carrier densities in these quantum wells are gate-tunable and electron mobilities up to 350,000 $\rm{cm^2(Vs)^{-1}}$ are extracted from magneto-transport measurements. Furthermore, from temperature-dependent magneto-resistance measurements, we extract an effective mass of 0.02 $m_0$ and find a Zeeman splitting compatible with the expected band edge g-factor. 
\end{abstract}

\maketitle

%


InSb elicits special interest in electronic\cite{ashley1995,orr2007}, electro-optical\cite{chen2005}, and spintronic\cite{zutic2004} applications due to its unique and extreme properties compared to other binary III-V compound semiconductors. Apart from the small band gap and electron effective mass, InSb is considered a candidate for the fabrication of topological quantum devices owing to its strong Rashba spin-orbit interaction (SOI)\cite{leontiadou2011,kallaher2010,khodaparast2004} and its intrinsic giant band edge g-factor of $|g|\sim51$\cite{qu2016,gilbertson2009}. As proposed by Y. Oreg\textit{ et al}.\cite{oreg2010}, a topological superconducting phase can be induced in a one-dimensional semiconductor with strong Rashba SOI in a Zeeman field by the coupling to an s-wave superconductor. Reports in this regard have been published for InAs\cite{Deng2016} and InSb\cite{Zhang2018} nanowire-based Majorana devices. In contrast to nanowires, two-dimensional electron gas (2DEG) systems are far more versatile for topological applications. Various types of scalable superconductor-semiconductor hybrid devices have been proposed\cite{Riwar2016,Stern2019,Peng2016} and experimental research in Al-InAs heterostructures\cite{Suominen2017,Fornieri2018,Whiticar2019} has hence followed. However, despite the superior intrinsic material properties, the progress on InSb 2DEGs is still hampered. Recently, there are developments in free-standing InSb nanostructures and their transport measurements, such as nanosails\cite{de2016twin} and nanosheets\cite{pan2016free,kang2018two,xue2019gate}. These layered InSb structures have advantages to achieve direct metal/superconductor contacts on them. Nevertheless, the research on InSb quantum wells (QWs) is still lacking due to the difficulties with heterostructure growth, although QWs have significant potential leading to high-quality devices\cite{yi2015gate,ke2019ballistic}. In QW systems, a thin barrier is required to induce superconductivity in the 2DEG through the proximity effect. However, the closeness of the 2DEG to the surface can limit the mobility of the carriers as a consequence. In this work, we present a quantum transport experiment, where the InSb 2DEG is close to the sample surface. Our magneto-transport measurements show that the 2DEGs still preserve a high mobility, even for the case of a QW with a barrier to the surface of only 5 nm. We also investigate other unique characteristics of InSb, such as the light electron effective mass $m^*$  and the large band edge g-factor. 

\begin{figure}
\includegraphics[width=0.5\textwidth]{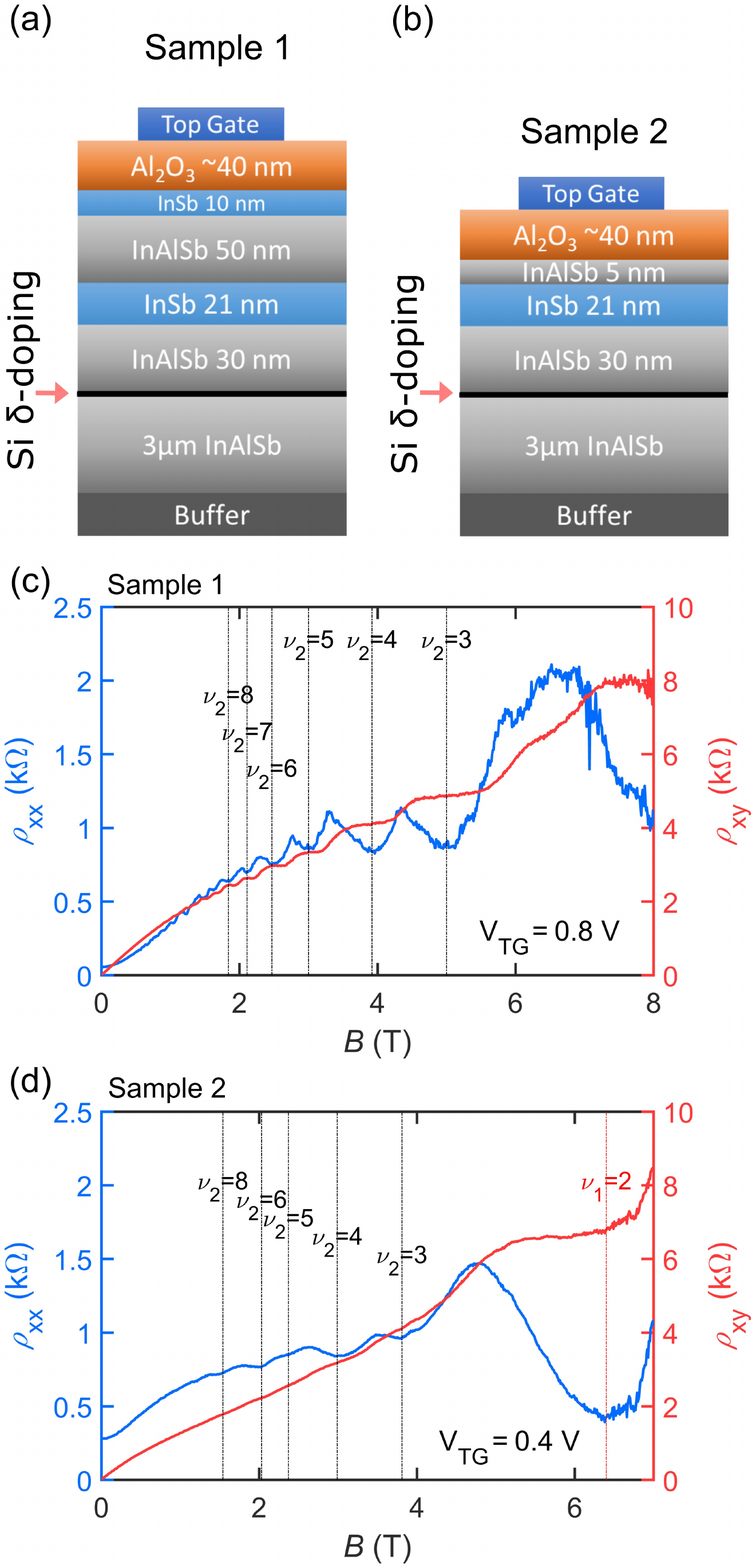}
\caption{\label{FIG.1} (a) Layer structure of sample 1. (b) Layer structure of sample 2. The $B$-dependence of $\rho_{xx}$ (blue) and $\rho_{xy}$ (red) of sample 1 and sample 2 are shown in (c) and (d), respectively. The filling factor of the electrons in the doping layer, $\nu_1$, and the filling factor of the electrons in the QW, $\nu_2$, are determined from the SdH data.}
\end{figure}

\begin{figure}
\includegraphics[width=0.5\textwidth]{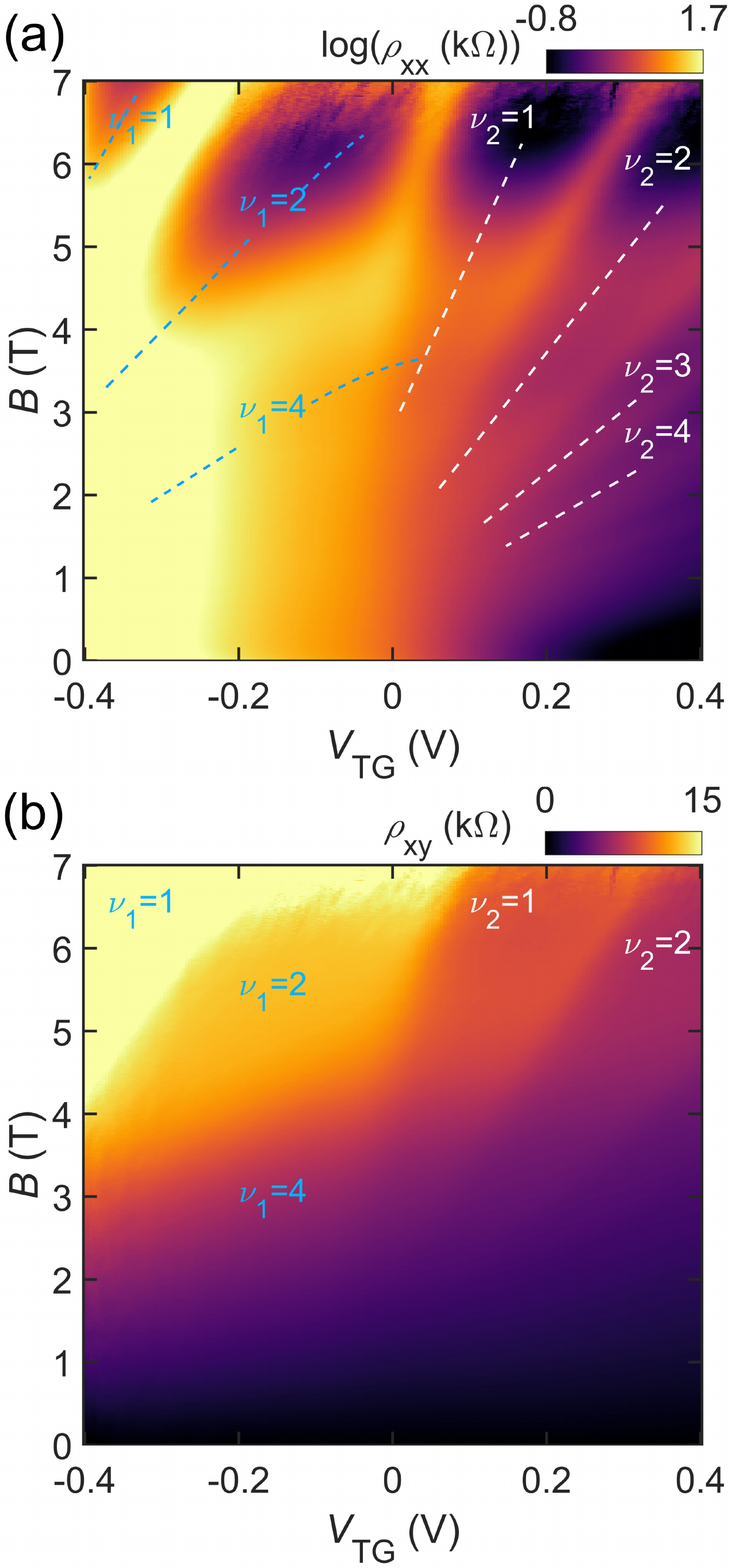}
\caption{\label{FIG.2}The detailed transport characterization of sample 2 at 1.3 K with $\rho_{xx}$ (a) and $\rho_{xy}$ (b) as functions of $V_{\rm{TG}}$ and $B$. The Landau fan diagrams and filling factors of the electrons in doping layer and QW layer are labeled with blue and white dashed lines respectively.}
\end{figure}

\begin{figure}
\includegraphics[width=0.45\textwidth]{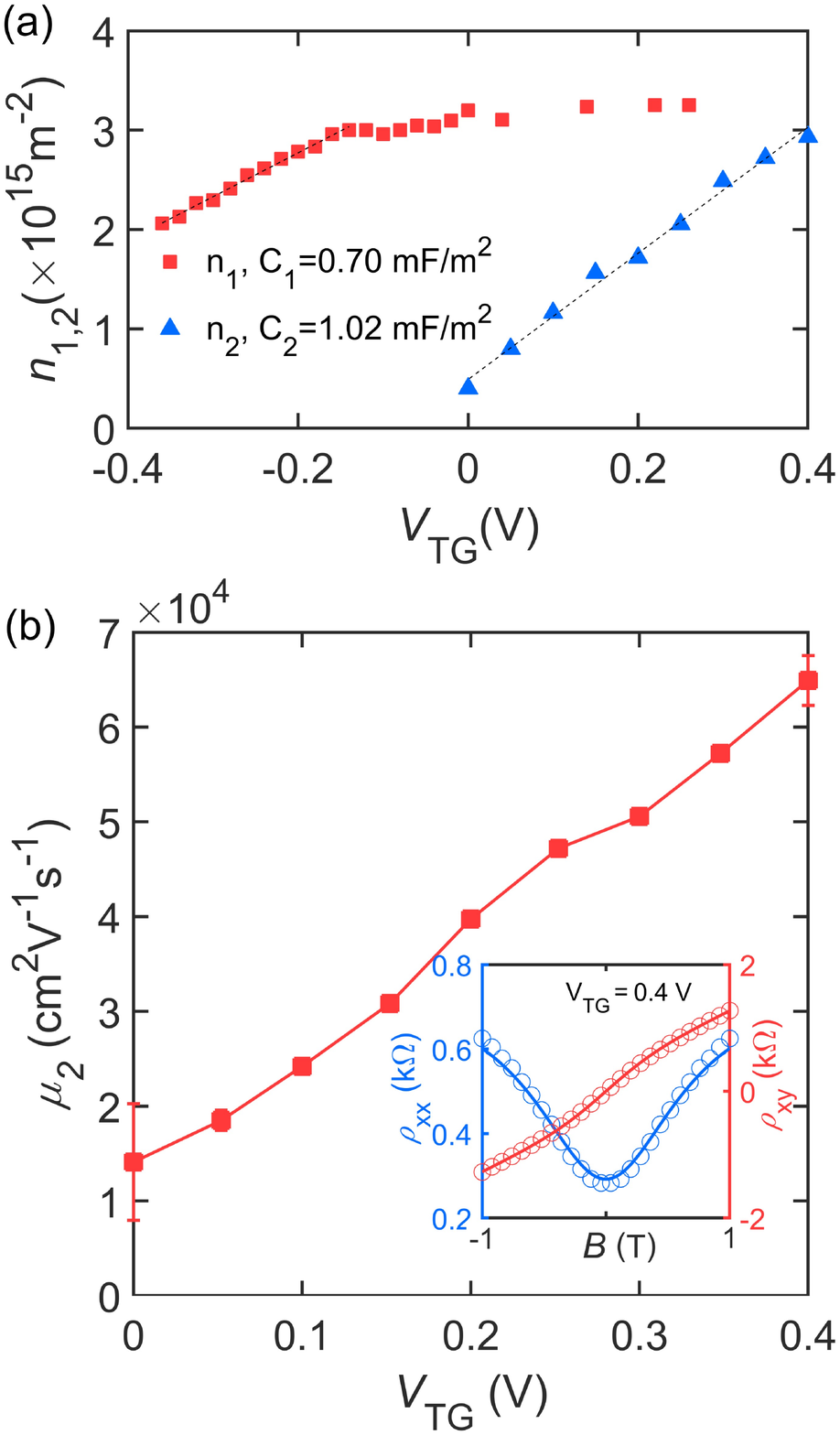}
\caption{\label{FIG.3}Analysis of the data shown in Fig. \ref{FIG.2}. (a) Carrier densities of the two conductive channels vs. $V_{\rm{TG}}$. (b) The mobility $\mu_2$ extracted from the two-band model vs. $V_{\rm{TG}}$. Insert: the data (circles) and fitting (lines) of $\rho_{xx}$ (red) and  $\rho_{xy}$ (blue) vs. $B$ when $V_{\rm{TG}}$=0.4 V. }
\end{figure}
\begin{figure}
\includegraphics[width=0.45\textwidth]{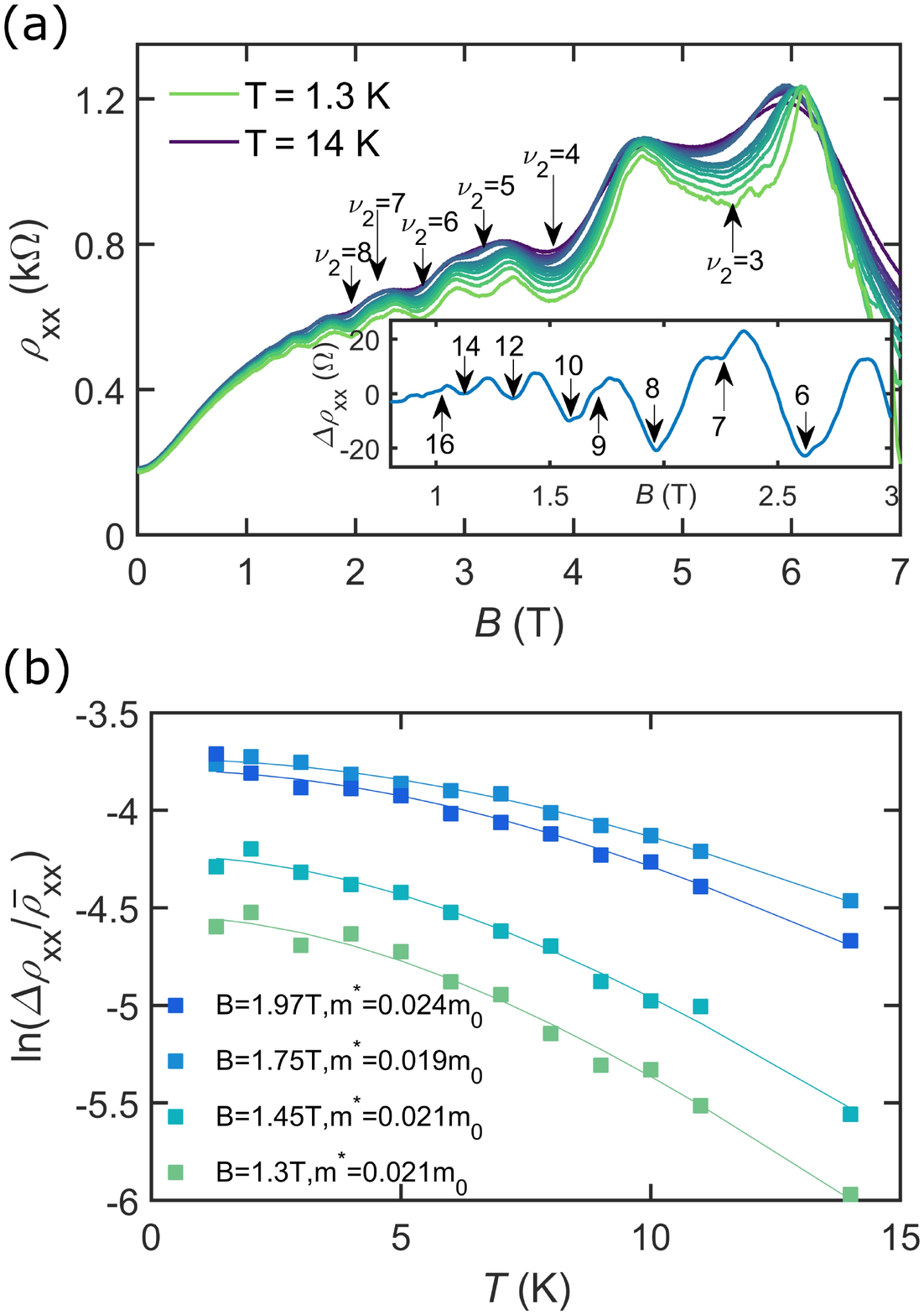}
\caption{\label{FIG.4} Effective mass measurement and the Zeeman splitting of sample 2. (a) Temperature-dependence of SdH oscillations with $n_2=3.8\times10^{15}$ $\rm{m^{-2}}$. Insert: $\Delta\rho_{xx}$ in a small magnetic field measured at 1.3 K. The Zeeman splitting happens at around $\nu_2=9$. (b) Dingle factor fitting with different $B$. The squares are data and the lines are fitted curves.}
\end{figure}
We report the fabrication and characterization of three InSb QW samples doped on the substrate side, which are grown on (100) GaAs substrates by molecular beam epitaxy (MBE). Two schematic layer sequences are shown in Figs.~\ref{FIG.1} (a) and (b). The growth details introduced in Ref.\onlinecite{Lehner2018} are only briefly outlined here. To overcome the lattice mismatch between GaAs and InSb, an interfacial misfit GaSb buffer as well as an interlayer InAlSb buffer are employed. The total thickness of this buffer system amounts to roughly 3 $\rm{\mu}$m. The 21 nm-thin InSb QWs are then surrounded by $\rm{In_{0.9}Al_{0.1}Sb}$ confinement barriers, while n-type carriers are introduced to the active region by a Si $\delta$-doping layer incorporated 30 nm below the QW in the barrier on the substrate side. On top of the QW, the $\rm{In_{0.9}Al_{0.1}Sb}$ layer thicknesses for samples 1 and 2 are 50 nm and 5 nm, respectively. Sample 1 is the only sample entailing a 10 nm InSb capping layer, while for samples 2 and 3 (25 nm thick upper barrier) the structure ends with $\rm{In_{0.9}Al_{0.1}Sb}$ [see Fig.~\ref{FIG.1}(b)]. We adopt the asymmetric bottom-doping scheme to avoid screening of the top gate electric field by the doping layer. A self-consistent band structure simulation of these 3 samples are introduced in the supplementary materiel.

Using wet chemical etching, standard Hall bar structures are defined with an etch depth of more than 120 nm, which is thus deeper than the Si-doping layer. Hall bar samples 1 and 2 have lateral dimensions of $50\times25$ $\rm{\mu m^2}$ (contact separation$\times$width) while sample 3 is $10\times4$ $\rm{\mu m^2}$ in size. Layers of Ge/Ni/Au evaporated on the contact areas of the samples after an Ar sputtering process provide Ohmic contacts to the 2DEG without the necessity of annealing\cite{Goel2005}. The samples are coated with a 40 nm-thick aluminium oxide (ALO) dielectric layer using atomic layer deposition at a temperature of 150 $^{\circ}$C. Finally, Ti/Au top gates covering the Hall bars are evaporated using electron beam evaporation. As a comparison, van der Pauw geometry samples without chemical etching and ALO are fabricated from the same wafers of the Hall bar samples. Their magneto-transport measurement is introduced in the supplementary material.  

The magneto-transport characterization is performed with standard low frequency (12 Hz) lock-in techniques at a temperature of 1.3 K. Figs.~\ref{FIG.1} (c) and (d) show the dependence of the longitudinal and transverse resistivities $\rho_{xx}$ and $\rho_{xy}$ of samples 1 and 2 in a magnetic field $B$ applied normal to the QW plane, where the top gate voltage $V_{TG}$ are 0.8 V and 0.4 V respectively. Shubinikov-de Haas oscillations in $\rho_{xx}$ and plateaus in $\rho_{xy}$ can be seen. In addition, the positive magneto-resistance and the nonlinear Hall resistance found in both samples at fields below about 4 T imply the existence of parallel conducting channels with distinct mobilities. Thus, the plateaus in $\rho_{xy}$ are not quantized at the expected values of the single-subband quantum Hall effect. The origin of the parallel channels is discussed below.

In the following, we describe the properties of sample 2 in detail, as it comprises the thinnest barrier. It is, therefore, most interesting with regard to a superconducting proximity effect induced by a superconducting contact. Details of samples 1 and 3 are found in the supplementary material. Figs.~\ref{FIG.2} (a) and (b) show $\rho_{xx}$ and $\rho_{xy}$ for sample 2 as a function of the top gate voltage $V_{\rm{TG}}$ and the magnetic field B, respectively. Fig.~\ref{FIG.2}(a) reveals two Landau fan diagrams, the first of which appears at low $V_{\rm{TG}}$ and is marked with blue dashed lines, while the second Landau fan diagram at high $V_{\rm{TG}}$ is marked with white dashed lines. The double-fan structure confirms the presence of two parallel conducting channels in the heterostructure. We extract their carrier densities $n_1$ and $n_2$ from the $1/B$ periodicity of the SdH oscillations. As shown in Fig.~\ref{FIG.3}(a), at lower $V_{\rm{TG}}$ only the first channel is populated and the density $n_1$ increases linearly with increasing $V_{\rm{TG}}$. When $V\rm{_{TG} >} -0.1$ $\rm{V}$, the increase in $n_1$ saturates while the second channel gets populated and  $n_2$ increases linearly instead. The gate capacitances of the first and the second channels are estimated to be $C_1=0.7$ $\rm{mF/m^2}$ and  $C_2=1.02$ $\rm{mF/m^2}$, respectively. We attribute $n_1$ to carriers in the Si-doping layer and $n_2$ to carriers in the QW. The calculated capacitances are within a factor of two of what is expected by considering the layer thicknesses and dielectric constants. The saturation of $n_1$ is due to screening of the gate electric field by the electrons populating the QW. 

A two-band Drude model allows us to estimate the mobilities $\mu_1$ and $\mu_2$ of the Si and the QW layer electrons. Using $n_1$ and $n_2$ obtained from the SdH oscillations, only the two mobilities remain as fitting parameters. As shown in Fig.~\ref{FIG.3}(b), the mobility $\mu_2$ increases with the increase of $V_{\rm{TG}}$. The insert of Fig.~\ref{FIG.3}(b) shows data (circles) and fitted curves (lines) of the low field $\rho_{xx}$  and $\rho_{xy}$ of sample 2 at a gate voltage $V_{\rm{TG}}=0.4$ $\rm{V}$ in a small magnetic field range. With $n_1=3\times10^{15}$ $\rm{m^{-2}}$ and $n_2=3\times10^{15}$ $\rm{m^{-2}}$, we find that mobilities are $\mu_1=7,500$ $\rm{cm^2 (Vs)^{-1}}$  and $\mu_2=67,000$ $\rm{cm^2(Vs)^{-1}}$, respectively. Phenomenologically, samples 1 and 3 behave similarly and their densities and mobilities are listed in Table 1. Especially, compared with previous related publications\cite{gilbertson2009,pooley2010quantum,gouider2010terahertz,leontiadou2011,uddin2012characterization,yi2015gate,ke2019ballistic,mlack2019plane}, sample 1 still holds similar or higher mobility of the 2DEG with comparable or thinner barrier thickness.  

From temperature-dependent SdH oscillations the effective mass of the electrons in the InSb QWs can be determined. Fig.~\ref{FIG.4}(a) shows the corresponding measurements for sample 2 with $n_2=3.8\times10^{15}$ $\rm{m^{-2}}$ determined from the $1/B$-periodicity. The oscillations of the resistivity $\Delta\rho_{xx}$ are obtained from subtracting the smooth background of the magneto-resistance $\bar{\rho}_{xx}$\cite{Habib2009}. Fig. \ref{FIG.4}(b) shows fits of the Dingle factor\cite{Ihn} to $\rm{ln}(\rho_{xx}/\bar{\rho}_{xx})$. The obtained effective mass is  $m^*\approx0.019$ $m_0$, where $m_0$ is the electron mass in vacuum. Using the same method, we find that the effective mass is density-independent within the range between $1.8\times10^{15}$ $\rm{m^{-2}}$ and $3.8\times10^{15}$ $\rm{m^{-2}}$. This result consist with the recent work by Ke \textit{et al}.\cite{ke2019ballistic}.

The spin-splitting of Landau levels is observed at magnetic fields 2 T, as shown in the inset of  Fig.~\ref{FIG.4}(a). With increasing $B$, the integer filling factor sequence changes from even to even and odd numbers. The magnetic field value beyond which the spin-splitting is resolved in the experiment is consistent with the band edge g-factor values determined in similar InSb QWs\cite{Lehner2018} and the works in InSb nanoconstrictions\cite{qu2016,nilsson2009giant,Nilsson2010correlation,ke2019ballistic}. A very rough estimate of the band edge g-factor for our device is found in the supplementary material. 

\begin{table}
\caption{\label{tab:Table1}Summary of carrier densities, mobilities and effective masses of all the three samples in this work. }
\begin{ruledtabular}
\begin{tabular}{cccc}
Properties&Sample 1&Sample 2&Sample 3\\
\hline
Upper barrier thickness&50 nm&5 nm&25 nm\\
$\mu_{2(max)}$ $\rm({cm^{2}(Vs)^{-1}})$&350,000& 67,000 & 160,000 \\
$n_2\rm{(\times10^{15}m^{-2})}$&0 - 3.5&0 - 3&0 - 3\\
$m^*$&0.020 $m_0$&0.019$\pm0.02$ $m_0$&-
\end{tabular}
\end{ruledtabular}
\end{table}
In summary, we have presented an InSb QW hetreostructure with inverted doping and ultra-thin InAlSb barriers to the sample surface. Using a standard Hall bar geometry, we performed magneto-transport measurements and found that the InSb QWs still show tunable densities and high mobilities, despite the small barrier thickness. The parallel conducting channel induced by the Si doping found in the investigated samples can be eliminated by reducing the doping concentration in future devices. We determined the effective in-plane mass of electrons in the InSb QWs to be 0.019 $m_0$ and estimated a large band edge g-factor close to the intrinsic bulk value. We anticipate that this work paves the way for realizing in-plane InSb devices with a superconducting top gate in the future, where Andreev reflection and 2D Majorana physics may be investigated.

~\\
\noindent{\textbf{SUPPLEMENTARY MATERIAL}}
~\\
See supplementary materials for the band edge g-factor estimation, the self-consistent band structure simulation, and the magneto-transport measurements of sample 1, sample 3 and the van der Pauw geometry samples.


\begin{acknowledgments}
This work was supported by the Swiss National Science Foundation through the National Center of Competence in Research (NCCR) Quantum Science and Technology. 
\end{acknowledgments}

\appendix

\nocite{*}
\bibliography{reference}

\newpage

%

\end{document}